# Surface Reconstructions in Molecular Beam Epitaxy of SrTiO$_3$

**Adam P. Kajdos and Susanne Stemmer**

Materials Department, University of California, Santa Barbara, California, 93106-5050, USA

**Abstract**

We show that reflection high-energy electron diffraction (RHEED) can be used as a highly sensitive tool to track surface and resulting film stoichiometry in adsorption-limited molecular beam epitaxy of (001) SrTiO$_3$ thin films. Even under growth conditions that yield films with a lattice parameter that is identical to that of stoichiometric bulk crystals within the detection limit of high-resolution x-ray diffraction (XRD), changes in surface reconstruction occur from (1×1) to (2×1) to c(4×4) as the equivalent beam pressure of the Ti metalorganic source is increased. These surface reconstructions are correlated with a shift from mixed SrO/TiO$_2$ termination to pure TiO$_2$ termination. The crossover to TiO$_2$ surface termination is also apparent in a phase shift in RHEED oscillations observed at the beginning of growth. Comparison with prior results for carrier mobilities of doped films shows that the best films are grown under conditions of a TiO$_2$-saturated surface [c(4×4) reconstruction] within the XRD growth window.

SrTiO$_3$ has long been investigated as a bulk material for its remarkable dielectric and superconducting properties [1,2]. More recently, it is also rapidly emerging as a building block for functional oxide interfaces, which can exhibit a wide range of emergent phenomena, including two-dimensional electron gases, magnetism, and superconductivity [3,4]. SrTiO$_3$ surface termination and reconstructions are important in determining interface and surface properties, such as interfacial conductivity [5], catalytic activity [6], and film growth [7]. Furthermore, reconstructions of a growing film's surface also provide important information about growth modes in high vacuum deposition techniques. For example, in molecular beam epitaxy (MBE) of III-V semiconductors, in-situ reflection high-energy electron diffraction (RHEED) studies of surface reconstructions are used to monitor the surface stoichiometry of the growing film [8-10]. Stoichiometry control of complex oxide films, such as SrTiO$_3$, is substantially more challenging, due to absence of a wide growth window in most deposition approaches, and the lack of physical characterization techniques that can characterize film stoichiometry with an accuracy better than ~ 1% [11]. A primary tool to characterize film stoichiometry is ex-situ high-resolution x-ray diffraction (XRD). For example, for SrTiO$_3$ it has been shown that both Ti and Sr excess give rise to an expansion of the lattice parameter relative to the stoichiometric value of 3.905 Å [12-15].

A growth window can be achieved in MBE of SrTiO$_3$ if one component, Ti, is supplied via a volatile, metal-organic precursor, such as titanium (IV) tetra-isopropoxide (TTIP) [15]. Within this growth window, a range of TTIP/Sr flux ratios result in films that have the bulk lattice parameter [15]. Using XRD for stoichiometry optimization is limited to deviations from stoichiometry that are sufficiently large to produce detectable peak shifts. This criterion may not be sufficient for electron-device quality films, which require stoichiometry control on the order of 0.1% or better. RHEED oscillations have been used in prior studies of MBE growth of SrTiO$_3$, but only to distinguish conditions that are also (non)stoichiometric according to XRD lattice parameter measurements [11,16,17]. SrTiO$_3$ surface reconstructions have been studied extensively by surface science techniques, and a good understanding exists of the relationship between specific reconstructions and the surface termination [18-22]. Here we show that surface reconstructions observed in RHEED can serve as a sensitive measure of SrTiO$_3$ film stoichiometry. In particular, changes in the reconstruction and changes in the RHEED intensity oscillations take place even within the growth window established by XRD. We show that adjusting growth parameters to yield a specific surface reconstruction serves as a more sensitive tool to optimize film stoichiometry than XRD alone, and is critical for obtaining the highest performance films.

SrTiO$_3$ thin films were grown on 10×10×0.5 mm$^3$ (001)SrTiO$_3$ single crystal substrates (MTI corporation) in a VEECO GEN930 MBE reactor (base pressure < 10$^{-9}$ Torr), using a similar process as described in Ref. [14]. Sr (99.99% purity, Sigma Aldrich) was supplied via a solid source effusion cell. Titanium was supplied via a low-temperature gas source using the metal-organic precursor titanium (IV) tetra-isopropoxide (99.999% purity, Sigma Aldrich). The Sr beam-equivalent pressure (BEP) was held constant at 5 × 10$^{-8}$ Torr while varying the TTIP BEP to achieve various TTIP/Sr ratios. Oxygen plasma was supplied via an RF plasma source operating at 250 W with a background pressure of 4 × 10$^{-6}$ Torr measured at the beam flux ionization gauge during growth. Substrate temperatures were measured using an Ircon Modline 3 optical pyrometer. All films are between 60 and 100 nm thick and were both inside and outside the XRD growth window. RHEED images were captured with a Staib Instruments RHEED system using a 14 kV accelerating voltage. The RHEED intensity of 00 and 10 reflections were measured simultaneously at the beginning of growth. The out-of-plane lattice parameters of the SrTiO$_3$ thin films were determined from high-resolution 2θ-ω x-ray diffraction (XRD) scans in the vicinity of the SrTiO$_3$ 002 reflection using a Philips X'Pert Panalytical MRD Pro Thin Film Diffractometer.

Figure 1 shows RHEED patterns captured after growth for films grown at different TTIP/Sr BEP ratios at a substrate temperature of 810 °C. Except for the sample grown at TTIP/Sr = 60.0, which was determined to be Ti-rich, all films lie within the XRD growth window. The patterns taken along the <100> and <110> azimuths are four-fold symmetric in all cases. The RHEED image taken for a film grown with TTIP/Sr = 37.5 indicates a (1×1) unreconstructed surface. The emergence of a ½-order reflection only along the <100> azimuths in the film grown with TTIP/Sr = 39.5 is indicative of mixed (1×2) and (2×1) reconstructed lattice domains. At TTIP/Sr = 45.0, the ½-order reflections along the [100] azimuth are more pronounced, and ¼-order reflections are apparent along the [110] azimuth, consistent with a c(4×4) reconstruction. The c(4×4) reconstruction persists up to a TTIP/Sr ratio of 60.0 (Ti-rich conditions, see Fig. 2), though clearly-defined chevrons along the <110> azimuths and dim ¼-order reflections along <100> azimuths are also observed at TTIP/Sr = 60.0. Both features are very similar to what is observed for mixed (1×4) and (4×1) anatase TiO$_2$ (001) domains grown epitaxially on SrTiO$_3$ (001) [23].

Figure 2 shows a map of the growth conditions that result in different reconstructed surfaces at growth temperatures of 810 °C and 850 °C, respectively, as well as the XRD growth window at each temperature as determined from measured film lattice parameters. Figure 3 shows RHEED intensity oscillations obtained at the beginning of growth for samples grown at 810 °C at various TTIP/Sr ratios. RHEED intensity oscillations for samples grown at 850 °C were either non-existent or decayed after no more than 6 oscillation periods, due to the transition from layer-by-layer to step-flow growth as a result of the higher growth temperature [14,15].

We first discuss the observed surface reconstructions seen in Figs. 1 and 2 in terms of their stoichiometry. In particular, these reconstructions indicate a systematic change in $TiO_2$ coverage of the growing surface. Previous scanning tunnel microscopy studies of bulk (001) $SrTiO_3$ surfaces have shown that a (1×1) reconstruction, observed here at the Sr-rich side of the growth window, are comprised of approximately equal fraction of SrO- and $TiO_2$-terminated sites [19,24]. The onset of a multi-domain (1×2) and (2×1) surfaces at higher TTIP/Sr values corresponds to expanding regions of coherent $TiO_2$ coverage, with uniformly distributed SrO-terminated sites [24]. Higher-order reconstructions, such as the c(4×4) observed here, or c(2×4) and c(2×6) observed by others, have been associated with purely $TiO_2$-terminated (001) $SrTiO_3$ surfaces [19,20,24]. Thus an increasing TTIP/Sr flux ratio during growth is closely tracked by the surface composition as exhibited by the different reconstructions. Most importantly, the boundary between the multi-domain (2×1) and the c(4×4) growth regimes marks the relative TTIP flux at which the growing surface is saturated with $TiO_2$. The shift of this boundary to higher TTIP/Sr at 850 °C compared to 810 °C is consistent with a higher desorption rate for TTIP at higher growth temperatures [15,25].

For growth conditions far outside the XRD growth window, the observed surface reconstructions are similar to those reported in previous studies of (001)$SrTiO_3$ surfaces. In the instance of substantial $TiO_2$ excess, either a weak (2×1) reconstruction (Fig. 2) or (1×1) surface (not shown) is typically observed, consistent with previous reports from both solid-source MBE [16,17] and hybrid MBE growth of $SrTiO_3$ (001) [14], and the diffraction pattern becomes increasingly diffuse, most likely due to amorphous excess $TiO_x$ [26]. In the SrO-excess regime, a c(2×2) is often observed, consistent with reports on SrO-rich surfaces [16,17]. These features are thus useful for determining the surface stoichiometry when growing outside the growth window, but such films are likely of limited practical use.

The changing character of the surface reconstruction with increasing TTIP flux suggest a substantial variation in the surface stoichiometry *within the XRD growth window* that likely reflects the growing film's stoichiometry. From Fig. 2 one concludes that for growth conditions within the growth window, the number of defects from non-stoichiometry incorporated into the bulk of the film is low enough such that the film and substrate lattice parameters are equal, despite these variations in surface stoichiometry. This conclusion is, however, dependent on the sensitivity of the lattice parameter to small deviations from stoichiometry, which is further complicated by substantial XRD peak overlap in homoepitaxy (i.e. no clearly defined separate film peak). In the (1×1) growth regime, the comparable fractions of SrO and $TiO_2$ surface terminations may result in a sufficiently low incorporation of defects to maintain film/substrate XRD peak overlap and thus fall within the XRD growth window. However, as a substantial portion of the surface is SrO-terminated and SrO has a low vapor pressure at typical growth temperatures, some defects may still be incorporated from SrO accumulation over time. Saturating the surface with volatile TTIP, which results in $TiO_2$ terminated surface reconstructions, precludes accumulation of defects by bringing the growth into a truly adsorption-controlled regime. The $TiO_2$ terminated surface is exposed to nonvolatile SrO, while excess TTIP desorbs to prevent excess $TiO_2$ accumulation. The higher-index c(4×4) reconstruction is a clear signature of this $TiO_2$ saturation. Although it is still present in excess $TiO_2$ conditions, outside the XRD growth window, the appearance of ¼-order reflections along <100>, combined with the dimming of ¼-order reflections and the appearance of chevrons along

<110>, for TTIP/Sr = 60.0 (Fig. 1) suggests faceted anatase-surface-like $TiO_2$ domains [23] coexisting with c(4×4) $TiO_2$-saturated domains. These features can be used as a signature of Ti-excess conditions. The growth conditions that give optimal stoichiometry can be identified by the overlap of the XRD growth window with the c(4×4) regime. La-doped $SrTiO_3$ films with the highest recorded electron mobilities for unstrained films, exceeding 50,000 $cm^2V^{-1}s^{-1}$ were grown in this regime [27].

The boundary between growth conditions yielding mixed termination and $TiO_2$ saturation is also apparent in the RHEED intensity oscillations at the beginning of growth, shown in Fig. 3. For TTIP/Sr ≤ 42.5, within the growth window, an abrupt phase shift in the intensity oscillations after ~5 monolayers from the specular 00 reflection puts the 00 reflected intensities almost $\pi$ rad out-of-phase with the 10 reflected intensities. For TTIP/Sr > 42.5, the 00 reflected intensities remain 0 to $\pi/2$ rad out-of-phase with the 10 reflected intensities without any obvious, abrupt phase shift, even for TTIP/Sr values slightly outside the XRD window (e.g. TTIP/Sr = 55.4). The phase shift is only observed for the 00 reflected intensities, which do not require surface features to have lateral symmetry to produce diffraction features (e.g. defects, steps) [28]. Thus it is likely that this phase shift is associated with the growth dynamics involving the boundaries between SrO- and $TiO_2$-terminated regions during layer-by-layer growth below the TTIP/Sr ratio corresponding to full $TiO_2$ surface coverage. Determining the exact origin of this phase shift would require more detailed investigation of the dynamics of $SrTiO_3$ growth by hybrid MBE, but within the scope of this study, this empirical observation serves as another indicator of the onset of $TiO_2$ saturation.

In summary, we have shown that MBE in an adsorption-limited regime allows for detailed monitoring changes in surface reconstructions even within the XRD growth window, where film lattice parameters indicate stoichiometric films. Both surface reconstructions and RHEED intensity oscillations provide insight into the extent of $TiO_2$ coverage on the growing surface. This enables precise control of the surface and resulting film stoichiometry, with a greater sensitivity than XRD alone, and thus enables the growth of highly perfect, stoichiometric $SrTiO_3$.

## Acknowledgments

This work was supported by the UCSB MRL, which is supported by the MRSEC Program of the U.S. National Science Foundation under Award No. DMR-1121053. A.P.K. also acknowledges support from the U.S. National Science Foundation through a Graduate Research Fellowship (Grant no. DGE-1144085).

**Figure Captions**

**Figure 1:** RHEED recorded along the [100] and [110] azimuths after the growth of SrTiO$_3$ films grown at different values of TTIP/Sr with a substrate temperature of 810 °C.

**Figure 2 (color online):** Out-of-plane lattice parameters of SrTiO$_3$ films grown at 810 °C and 850 °C as a function of the TTIP/Sr ratio and the corresponding observed reconstructions at each growth condition. The dashed lines correspond to the bulk lattice parameter of cubic SrTiO$_3$. The field labeled (2×1) refers to a multi-domain (2×1)/(1×2) surface.

**Figure 3 (color online):** RHEED intensity oscillations measured along the [100] azimuth at the beginning of growth for SrTiO$_3$ films grown at various TTIP/Sr with a substrate temperature of 810 °C. Black arrows mark the approximate onset of each phase shift. The beginning of each growth involved compensation of a Sr source flux transient to maintain constant TTIP/Sr in the early stages of growth.

**Figure 1**

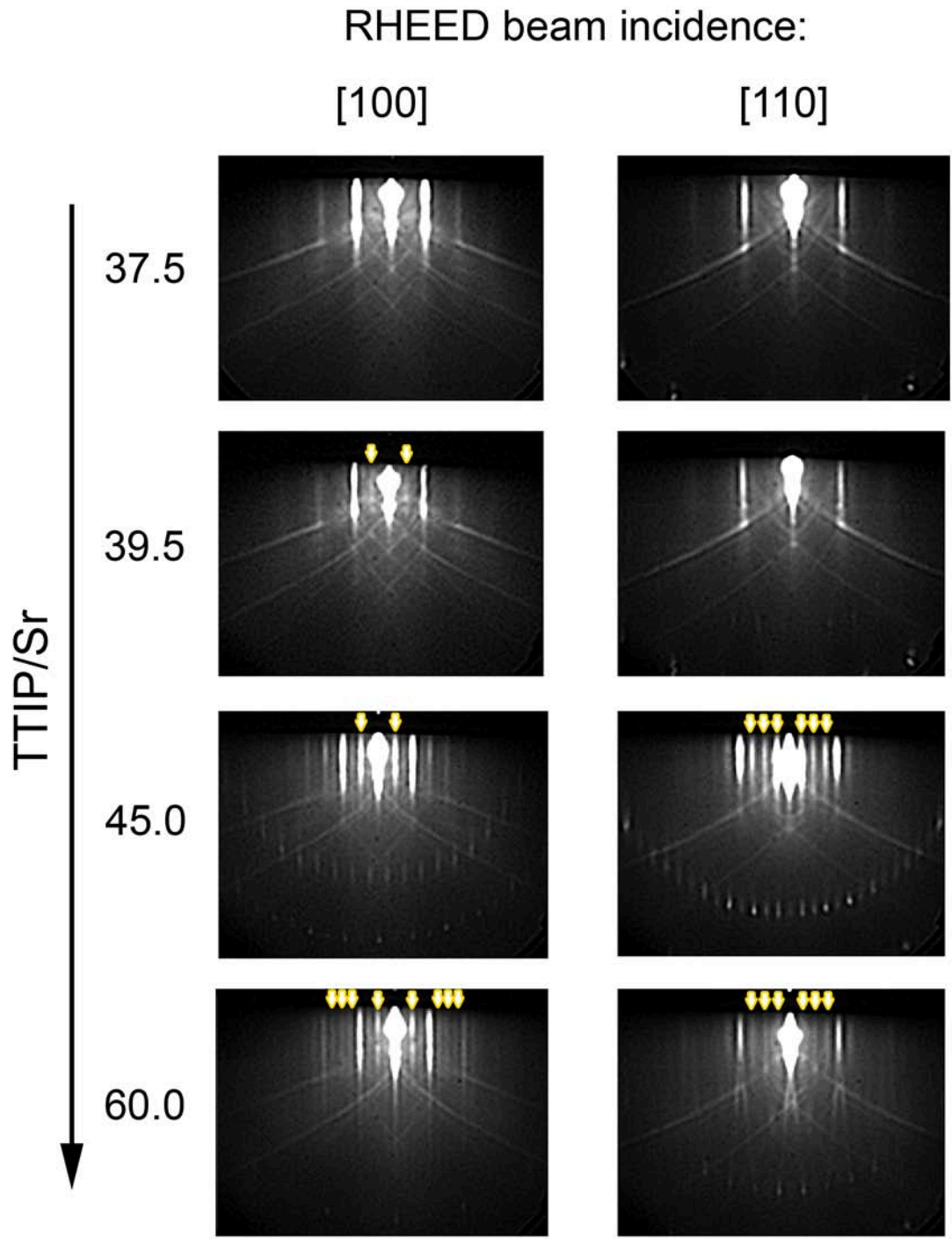

**Figure 2**

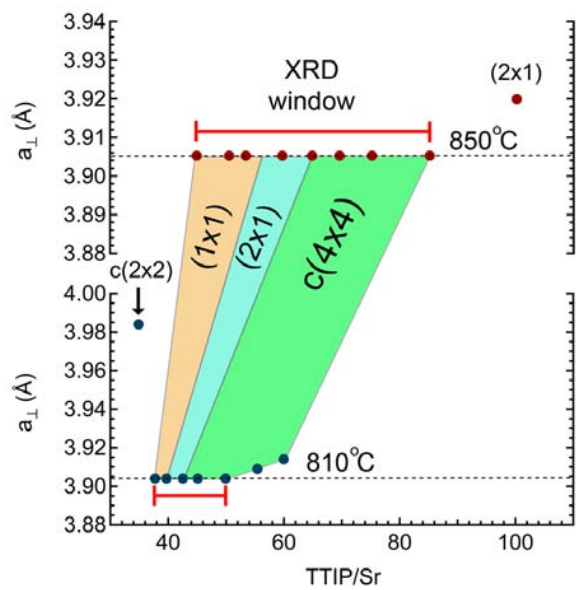

**Figure 3**

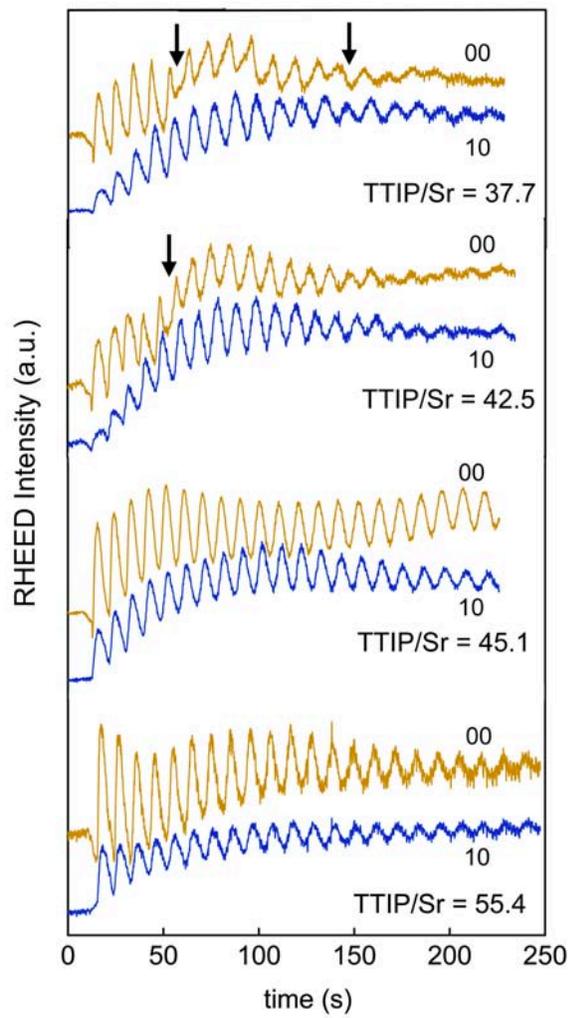